# DEEP IMAGING OF THE DOUBLE QUASAR 0957+561:
# NEW CONSTRAINTS ON $H_0$

H. Dahle[1,3], S.J. Maddox[2,4], and Per B. Lilje[1,5]

[1] *Institute of Theoretical Astrophysics, University of Oslo, P.O. Box 1029 Blindern,*
*N-0315 Oslo, Norway*

[2] *Royal Greenwich Observatory, Madingley Road, Cambridge CB3 0EZ, England*

## ABSTRACT

In this *Letter* we present new results from extremely deep, high-resolution images of the field around the double quasar QSO 0957+561. A possible gravitational arc system near the double quasar has recently been reported, which, if real, would set strong constraints on determinations of the Hubble constant from the time delay in the double quasar. We find that both the morphology and the colors of the claimed arc systems suggest that they are chance alignments of three and two different objects, and not gravitationally lensed arcs. Hence, the constraints on $H_0$-determinations from the arcs are not valid. Also, a small group of galaxies at $z = 0.5$ near the line-of-sight which was required to have a very large mass in the physically interesting arc models, is most likely insignificant.

From our deep images we are able to use weak lensing of faint background galaxies in the field to map the gravitational potential in the main cluster. This sets new constraints on determinations of $H_0$. We find that the Hubble constant is constrained to be less than $70\,\mathrm{km\,s^{-1}\,Mpc^{-1}}$, if the time delay between the two images of the QSO is equal to or larger than 1.1 years.

*Subject headings:* cosmology: distance scale — gravitational lensing

---

[3] E-mail address: hakon.dahle@astro.uio.no
[4] E-mail address: sjm@mail.ast.cam.ac.uk
[5] E-mail address: per.lilje@astro.uio.no

# 1. INTRODUCTION

The "double quasar", QSO 0957+561, discovered by Walsh, Carswell, & Weyman (1979), remains the best studied example of a gravitational lens system. As had already been shown by Refsdal (1966), a value for the Hubble constant which does not rely on the steps of the local distance ladder may be determined from the time delay between the different images of the same variable source in such gravitationally lensed systems. The QSO 0957+561 system is the only one where a fairly accurate, although not uncontested, value for the time delay has been obtained (e.g., Florentin-Nielsen 1984; Press, Rybicky, & Hewitt 1992; Pelt *et al.* 1994). However, as was shown first by Borgeest & Refsdal (1984) and later elaborated by e.g., Falco, Gorenstein, & Shapiro (1991) and Kochanek (1991), a determination of the Hubble constant will not only require a reliable value for the time delay between the two quasar images, but also a well-constrained model for the lensing mass distribution. The main problem is the complicated mass distribution of this lens, due to the fact that the lensing galaxy is near the center of a galaxy cluster which gives a significant contribution to the total lensing potential. As well as the cluster at $z = 0.36$ there is also another galaxy group at $z = 0.5$ centered on a galaxy 1'.5 from the lens center. Even in the simple models of Kochanek (1991) which ignore this group there are nine free parameters describing the shape and position of both the lensing galaxy and the surrounding cluster. Accurate VLBI measurements of the positions and flux ratios of the QSO images and their radio lobes (e.g., Gorenstein *et al.* 1988; Conner, Lehár, & Burke 1992) provide 5 constraints on the model. Further constraints are therefore necessary if one wants to use this system for a determination of the Hubble constant.

A possible strong constraint was reported by Bernstein, Tyson, & Kochanek (1993). They found a pair of elongated objects $\sim 20"$ from the lens center, which they interpreted as strongly lensed background galaxies. This interpretation relied solely on morphological evidence. The existence of these arcs restricted models for the lensing density distribution to just two classes:

1. Single-screen models in which the lensing galaxy has a very small mass, or

2. Two-screen models which need to include a large mass ($\sim 10^{14} M_\odot$) in the $z = 0.5$ group.

Class 1 models require unreasonably low values for the galaxy mass, and give very low estimates of $H_0$, but if class 2 models are correct, we would be forced to accept that a rather unassuming group of 4-5 galaxies has a total mass equal to that of a rich cluster of galaxies.

Another way of constraining the mass distribution is to measure the weak gravitational lensing effect of the cluster of galaxies by observing the distortion of the images of faint background galaxies, a technique first used by Tyson, Valdes, & Wenk (1990). From sufficiently deep images, this effect can be utilized to map the projected mass distribution of the cluster (Kaiser & Squires 1993). If it is possible to measure this effect in the field of the double quasar, it will set interesting constraints on the mass distribution model.

In this *Letter* we describe new observations of the QSO field which suggest that the possible arcs are in fact unlikely to be lensed images of background galaxies. We also use the distribution of faint background galaxies to estimate the distribution of mass in the lens, and hence set new constraints on $H_0$.

## 2. OBSERVATIONS AND DATA REDUCTIONS

We observed the field around QSO 0957+561 on the three nights of January 7–10, 1994 with the 2.56 meter Nordic Optical Telescope (NOT), using the IAC CCD camera. The camera was kindly provided by the Instituto de Astrofisica de Canarias and is equipped with a coated Thomson $1024 \times 1024$ pixel CCD chip which gives an image scale of 0".14 pixel$^{-1}$ at the NOT. We took a large number of frames, shifting the field by a few arc-seconds between each exposure. Each integration was typically 1200 s which ensured that sky was the dominant noise source. The total integration time was 24,600 s in the $V$-band and 27,000 s in the $I$-band. The images were bias subtracted with the IRAF package, which was also used for other data reductions. Flat-fields were made from the sky background in the science images by median averaging all images taken in each filter on the same night before they were registered. Before averaging, objects above a threshold of 3 times the standard deviation of the sky background were detected in each frame, and the pixels near each object were not used in the averaging. With these master flats, all frames in each color were flat-fielded and then registered and co-added using the standard IRAF sigma clipping technique. We thus end up with two very deep 2'×2' images in the $V$ and $I$ bands. The combined $I$ frame has a seeing of 0".76 (FWHM), and the combined $V$ frame has a seeing of 0".81. Object detection and photometry was done with the FOCAS software and was calibrated with standard stars in the NGC 7790 and NGC 4147 fields (Christian *et al.* 1985). The $2.5\sigma$ detection limit for the object catalogs are $V \approx 27$ and $I \approx 26$.

## 3. THE POSSIBLE ARC SYSTEM

Contour plots of a $12''.6 \times 14''.1$ field centered on the position of the possible arc system of Bernstein *et al.* (1993) from our images are shown in Figure 1 (the $V$-image in Figure 1a and the $I$-image in Figure 1b). The "arc system" of Bernstein *et al.* (1993) is labelled as A1 and A2. As shown in the Figure, A1 and A2 are resolved into several distinct clumps. Two substructures were found in A2 and three were found in A1 (here designated as A1a, A1b, etc). Four of the five arc substructures were detected in both $I$ and $V$, while A1c was only seen in the $V$ frame. All five substructures were classified as galaxies by FOCAS. Figure 1b shows a rather large gap between the two main clumps in A1: The two structures A1a and A1b are separated by about 4", and there is no convincing signal in the area between them. The arc candidate A2 seems convincingly resolved as well, although the separation between the components is smaller (1".8) than for A1.

The morphology of A1 and A2 suggests that these are not gravitationally lensed, stretched-out images of background galaxies, but rather separate objects which may be close, interacting galaxies or chance alignments of galaxies at completely different redshifts. We cannot, however, completely rule out the arc nature of these objects based only on morphology, as some real gravitational arcs are quite lumpy because of structure in the lensing potential or in the lensed galaxy (although the morphology of the lensed galaxy would have to be very strange if it should cause the observed structures).

To give a definite answer to this problem, color information is essential. If A1 and A2 are stretched-out images of a single galaxy, one would certainly not expect to see large color differences between the various parts of the arcs. Figure 2 shows a color–magnitude diagram for all detected objects in the field. The two quasar images are marked with crosses, the two brightest components of A1 with asterisks, and the two components of A2 with triangles. Photometry of the arcs (see Table 1) shows that the $V-I$ colors of the components in A2 differ by 1.5 magnitude, and the components of A1 by at least 1.4 magnitudes. Although the differing lines of sight could pass through regions of different reddening, these color differences strongly argue against the components being images of the same background galaxy.

Based on both photometric and morphological evidence, we conclude that A1 and A2 are unlikely to be gravitational arcs. This means that there is no evidence for a large mass in the $z = 0.5$ group, and that the constraints imposed by Bernstein *et al.* (1993) on determinations of $H_0$ from the time delay in the double quasar are invalid.

## 4. MAPPING THE GRAVITATIONAL POTENTIAL

The gravitational potential of the cluster of galaxies in front of QSO 0957+561 should distort images of background galaxies through weak gravitational lensing. Hence we can use the shapes and orientations of faint background galaxies to map the lensing mass. Our method will be described in detail in a future paper and we give only a brief overview here. The shape and orientation of each background galaxy can be considered as an estimate of the local shear caused by the lens. The unknown intrinsic ellipticity of each galaxy adds noise to the shear estimate, but averaging over a large number of background galaxies reduces the noise to a level where the gravitationally induced shear can be accurately determined. We selected 123 galaxies with $26.5 > V > 23.5$ and $V - I < 1.5$, and used the FOCAS measurements of the second moments to estimate the position angle and elongation for each image.

We used the method developed by Kaiser (1992) and Kaiser & Squires (1993) to invert the estimated shear field and recover a map of the projected cluster mass surface density. The noise in the map is larger if a higher resolution is used; we found that a Gaussian window function with $\sigma = 9''$ was a reasonable compromise for our data. The resulting map of surface mass density is shown in Fig 3. The significance of the features in the map were tested by varying the magnitude limits of the sample and by bootstrap sampling.

There is only one significant peak in the mass distribution, and this is centered on the main galaxy cluster, $6''$ from the lensing galaxy in the direction $\theta_c = 123°$ (using the notation of Kochanek where angles are measured from west, increasing towards north). Our bootstrap resampling technique shows that it has a signal-to-noise ratio larger than 3. Its position is $6''$ from the galaxy number or luminosity weighted position estimated from the Young *et al.* (1981) catalog. I.e., the center of the inferred mass peak is near the center of light, but the mass distribution seems to be elongated in a different direction than the light distribution. However, the shape of both the mass- and light distributions are quite poorly determined. There is no evidence for a significant peak at the position of the galaxy group at $z = 0.5$, which argues against this group being important in the lensing system. A further strong argument against the importance of the group is that we see no excess of faint galaxies around it.

## 5. CONSTRAINTS ON $H_0$

Both the rejection of A1 and A2 as gravitational arcs, and the apparent insignificance of the mass density near the galaxy group at $z = 0.5$ suggest that the simple cluster plus galaxy models of Kochanek (1991) provide a realistic description of the lens system. These models consist of an elliptical pseudo-isothermal cluster potential offset from an isothermal galaxy potential which is either elliptical (model 1) or has a quadrupole term which decreases $\propto 1/r^2$ (model 2). The models have 9 free parameters, and so the 5 VLBI constraints allow many solutions. The relative position of the galaxy and cluster potentials is important in constraining $H_0$, but was previously poorly determined. The position of the peak in our mass density map provides a reliable estimate of the cluster position, which removes two of the 4 free parameters, and so considerably restricts the allowed solutions.

Our estimate of the cluster position relative to the central galaxy is $r_c = 6''$ and $\theta_c = 123°$ (where the notation is as explained in § 4). From analyzing different subsamples of background galaxies, and by applying bootstrap resampling error analysis, we estimate the uncertainty in the position of the peak to be 3" in $r_c$ and $30°$ in $\theta_c$. Using Kochanek's models, we see from his Fig. 7 that the scaled time delay is restricted to lie between 1.5 and 7.0. Kochaneck's class 1 models have no solutions consistent with our estimates of $\theta_c$; only his class 2 models are allowed.

The major uncertainty in the estimates of the scaled time delay is from the uncertainty in the mass of the lensing galaxy. The values quoted above refer to galaxy velocity dispersions between 240 and 320 $km\,s^{-1}$ (Rhee [1991] measures $305 \pm 50\,km\,s^{-1}$). Recent determinations of the time delay give either values of approximately 410 days (e.g., Pelt *et al.* 1994) or approximately 536 days (e.g., Press *et al.* 1992). In combination with our results the former gives the restriction $15\,km\,s^{-1}Mpc^{-1} < H_0 < 69\,km\,s^{-1}Mpc^{-1}$ while the latter gives $11\,km\,s^{-1}Mpc^{-1} < H_0 < 52\,km\,s^{-1}Mpc^{-1}$. We conclude that the measured time delay in the double quasar QSO 0957+561 in combination with our determination of the gravitational mass center of its lensing cluster give a strong constraint that $H_0 < 70\,km\,s^{-1}Mpc^{-1}$.

We gratefully acknowledge the use of the Nordic Optical Telescope and thank the Instituto de Astrofisica de Canarias for giving us access to their CCD camera. We thank the NOT staff, and especially Anlaug Amanda Kaas, Anton Norup Sørensen, and Leif Festin, for their support during our observing run. This work has benefited from useful discussions with Sjur Refsdal and Mogens Hansen.

**TABLE 1**

Photometry of the objects in A1 and A2

| Object | $V$ | $I$ | $V - I$ |
|--------|------|------|--------|
| A1a | 26.4 | 24.5 | 1.9 |
| A1b | 25.9 | 24.5 | 1.4 |
| A1c | 25.7 | — | < 0.5 |
| A2a | 25.3 | 23.8 | 1.5 |
| A2b | 24.9 | 24.9 | 0.0 |

# REFERENCES

Bernstein, G.M., Tyson, J.A., & Kochanek, C.S. 1993, AJ, 105, 816

Borgeest, U. & Refsdal, S. 1984, A&A, 141, 318

Christian, C.A., Adams, M., Barnes, J.V., Butcher, H., Hayes, D.S., Mould, J.R.,
     & Siegel, M. 1985, PASP, 97, 363

Conner, S., Lehár, J., & Burke, B.F. 1992, ApJ, 387, L61

Falco, E.E., Gorenstein, M.V., & Shapiro, I.I. 1991, ApJ, 372, 364

Florentin-Nielsen, R. 1984, A&A, 138, L19

Gorenstein, M.V., Cohen, N.L., Shapiro, I.I., Rogers, A.E.E., Bonometti, R.J.,
     Falco, E.E., Bartel, N., & Marcaide, J.M. 1988, ApJ, 334, 42

Kaiser, N. 1992, ApJ, 388, 272

Kaiser, N. & Squires, G. 1993, ApJ, 404, 441

Kochanek, C.S. 1991, ApJ, 382, 58

Pelt, J., Hoff, W., Kayser, R., Refsdal, S., & Schramm, T. 1994, A&A, *in press*

Press, W.H., Rybicki, G.B., & Hewitt, J.N. 1992, ApJ, 385, 416

Refsdal, S. 1966, MNRAS, 132, 101

Rhee, G.F.R.N. 1991, Nature, 350, 211

Tyson, J.A., Valdes, F., & Wenk, R.A. 1990, ApJ, 349, L1

Walsh, D., Carswell, R.F., & Weyman, R.J. 1979, Nature, 279, 381

Young, P., Gunn, J.E., Kristian, J., Oke, J.B., & Westphal, J.A. 1981, ApJ, 244,
     736

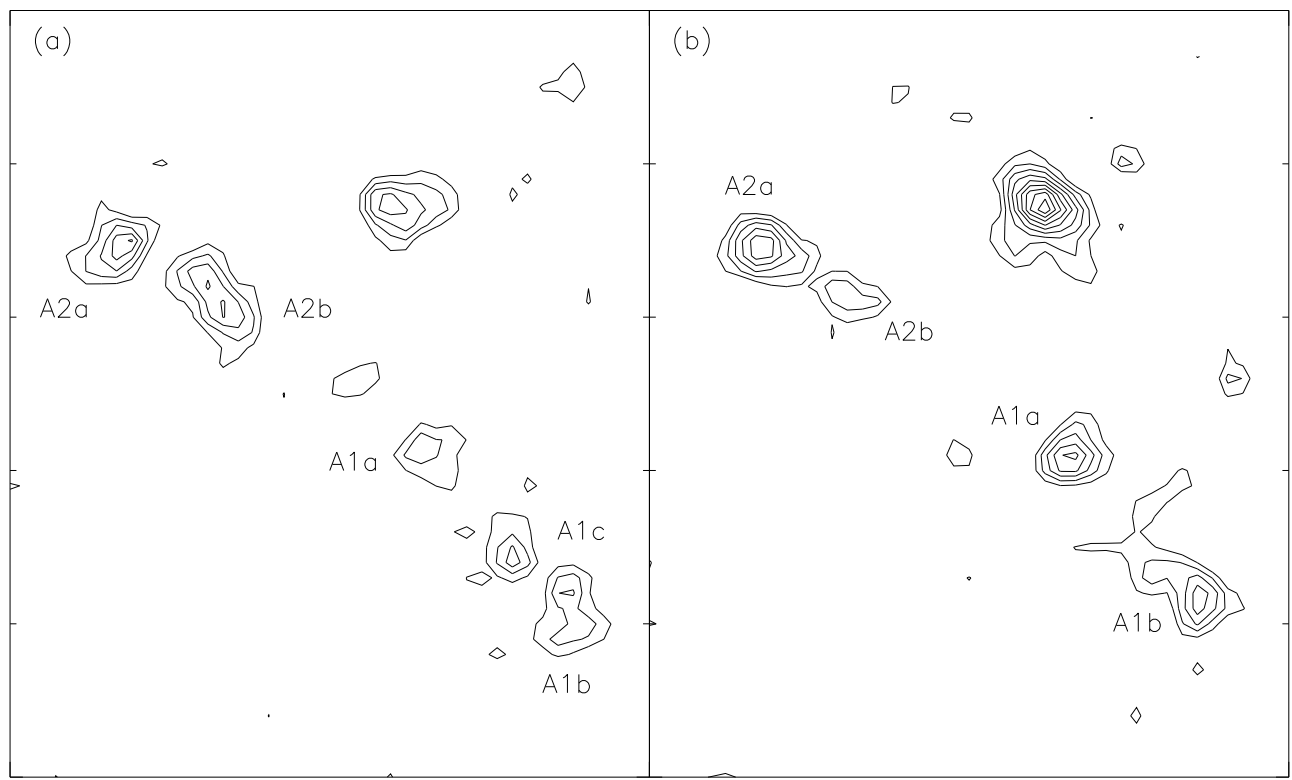

FIG. 1.—Contour maps of the $V$- (Fig. 1a) and $I$- (Fig. 1b) images in a $12''.6 \times$ $14''.1$ area centered on the proposed arc.

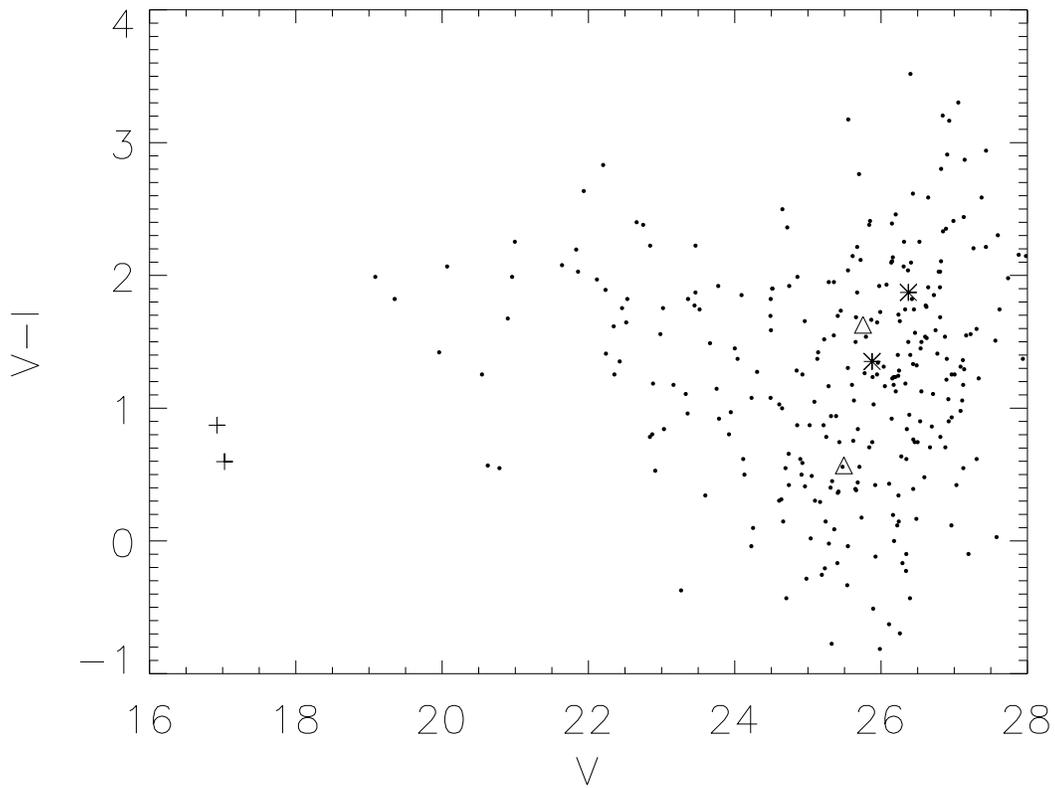

FIG. 2.—Color-magnitude diagram for objects in the 0957+561 field. The two quasar images are marked with crosses, and the four arc components with known colors are marked as asterisks (A1, a and b) and triangles (A2). The photometry of the quasar images is not performed in an optimal way, and the color difference between the two components is caused by the B image being merged with the redder lensing galaxy.

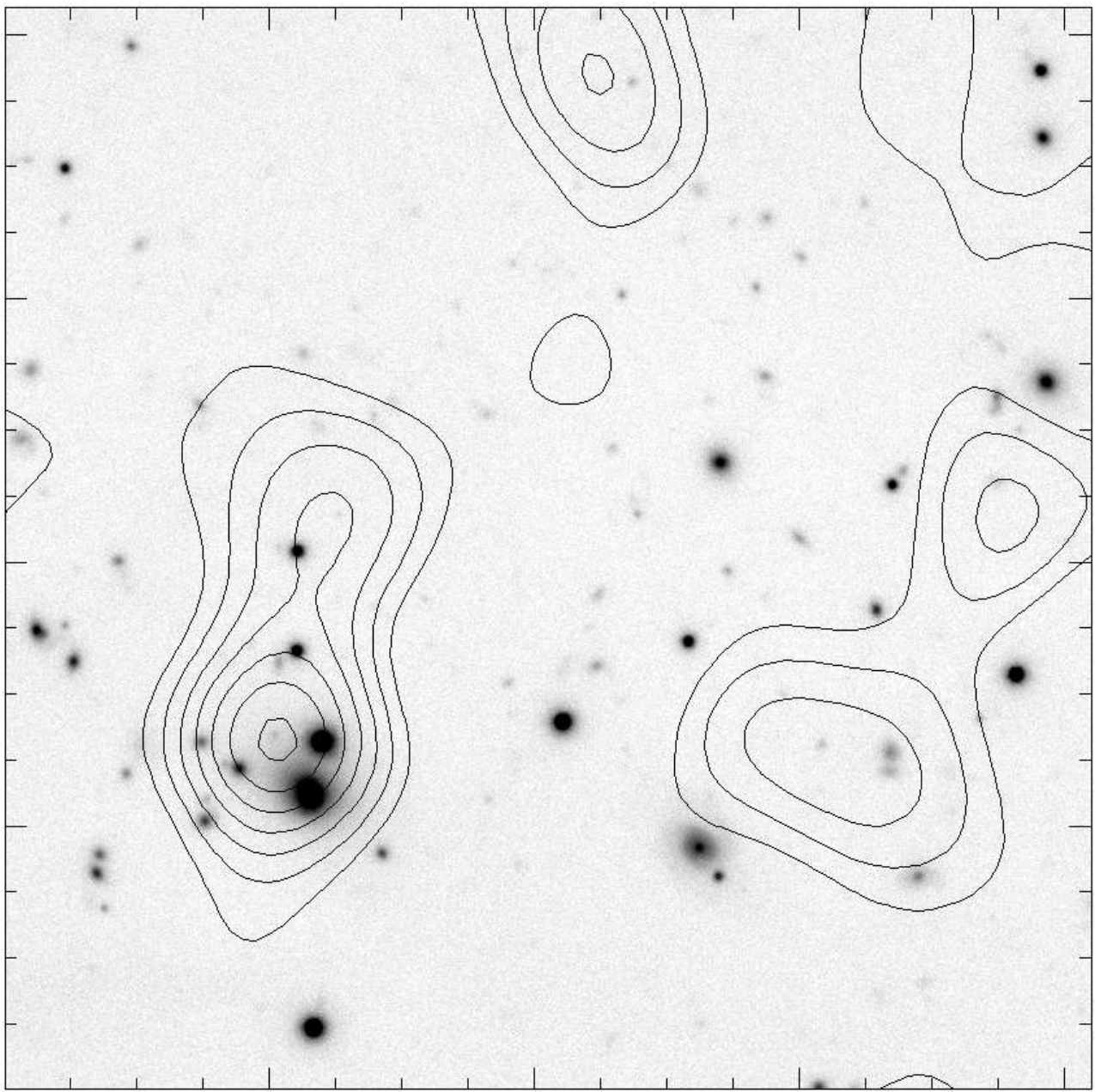

FIG. 3.—Contour map of the inferred projected mass density in the field around QSO 0957+561 overlaid our *I*-image. North is up and west to the right. The double

quasar consists of the two bright images in the SE quarter, the A-image is the northernmost, while the B-image together with the image of the lensing galaxy is the southernmost. The contour levels are in steps of 0.5 times the average rms noise in the mass-map, determined by bootstrap resampling (the noise varies substantially with position in the image). Only the projected mass peak near the double quasar is found to be significant by the bootstrap resampling.